\documentclass[a4paper]{ESASPCS13Style}
\usepackage{epsfig}

\newcommand{\cmcc}{cm$^{-3}$}

\newcommand{\ergps}{erg~s$^{-1}$}

\newcommand{\chan}{\textit{Chandra}}
\newcommand{\xmm}{\textit{XMM-Newton}}
\newcommand{\asca}{\textit{ASCA}}
\newcommand{\euve}{\textit{EUVE}}
\newcommand{\sax}{\textit{Beppo}SAX}

\begin{document}

\title{X-Rays of Stellar Coronae with Chandra and XMM-Newton}
\subtitle{Flares and Elemental Composition in Stellar Atmospheres}

\author{Marc Audard\inst{1}}
  
  \institute{Columbia Astrophysics Laboratory, Columbia University, Mail code 5247, 550 West 120th
  Street, New York, NY 10027, USA}

\maketitle 

\begin{abstract}

Observations of magnetically active stars with \chan\  and \xmm\  have deepened our knowledge 
of the physics of the atmospheres in late-type stars. In this review paper, I discuss two topics 
that have profited significantly from \chan\  and \xmm. Of particular interest, studies of the 
elemental composition of stellar coronae have taken advantage of the high spectral resolution available 
with the grating instruments on board these satellites. I summarize the status of our knowledge and 
discuss the elemental composition in a variety of stellar coronae. I also focus on the topic of 
flares in stellar coronae, and review the contributions made by \xmm\  and \chan\  to X-ray 
and multi-wavelength studies.

\keywords{Stars: abundances -- Stars: activity -- Stars: atmospheres -- Stars:
coronae -- Stars: flares -- X-rays: stars}
\end{abstract}

\section{Introduction}

X-ray astronomy of stars has been a rich and prosperous field for more than a quarter of
a century (see reviews by \cite{favata03,guedel04a}). Late-type stars display strong X-ray emission
as a signature of magnetic activity in their upper atmospheres (e.g., \cite{linsky85}). Stellar
coronae have been studied in detail with many extreme ultraviolet (EUV) and X-ray satellites, 
but the launches of both \chan\  and \xmm\  finally gave access to high-resolution 
spectroscopy in the X-ray regime with high sensitivity. Of particular interest, 
the He-like triplets have given us access to electron densities which are of crucial importance
to derive coronal volumes from emission measures (EMs). In addition, the elemental composition
of the stellar coronal plasma can be obtained without much spectral confusion and the complication
of radiative transfer for several abundant elements such as C, N, O, Ne, Mg, Si, S, Ar, Ca, and Fe. 

Whereas the Sun's proximity helps to study the physical processes in its upper atmosphere, the investigation
of X-ray coronae provides us with a wide range of physical conditions different from the Sun (e.g., mass,
radius, rotation period, luminosity). The lack of spatial resolution in stars
also helps us to obtain a global view of coronal physics in its most extreme conditions. Indeed, 
the solar corona does not sustain a corona with a dominant plasma temperature of a few tens of MK, 
but stars do.

This review summarizes the results obtained with the new \xmm\  and \chan\  X-ray satellites in the field of
stellar coronae. About five years after their launches in 1999, there is a vast pool of publications and
topics to review, too vast for merely 10 pages. Therefore, this review focuses on two specific topics, 
namely the elemental composition of stellar coronae (\S\ref{sec:compos}), and stellar flares 
(\S\ref{sec:flares}).
Although I will address various issues related to abundances and flares, completeness is not 
possible given the limited space. I will therfore discuss selected representative examples 
in more detail but touch upon others only in a cursory way.
A complementary review on the topics of densities and coronal structures can be found 
in these proceedings as well (\cite{ness05}). Previous reviews summarizing the view three years
after the launch of \chan\  and \xmm can be found elsewhere
(\cite{audard03a,linsky03}).

\section{Elemental Composition of Stellar Coronae}
\label{sec:compos}

\subsection{A brief introduction}

In the Sun, coronal abundances show a specific pattern in which elements with first ionization potentials
(FIP) $< 10$~eV are enriched by a factor of $\sim 4$ with respect to their photospheric abundances. Although
this so-called FIP effect can be observed in the full-disk emission of the Sun (\cite{laming95}), details
show, however, different FIP bias depending on where coronal abundances are
measured (e.g., fast vs. slow wind, active regions, coronal holes, etc; see, e.g., \cite{feldman00}).

Coronal abundance studies in stars in the pre-\chan\  and \xmm\  era were promising but still in an infant state.
Only the \euve\  and \asca\  satellites had sufficient spectral resolution to derive a few element 
abundances. The abundance of Fe in active stars was found to be deficient by factors of 5 to 10 relative to 
the solar photospheric abundance (e.g., \cite{drake96,schmitt96}). On the other hand, a solar-like FIP
effect was derived in inactive stars (\cite{drake97,laming99}). 
The advent of \chan\  and \xmm\  have given further insight into the elemental
composition in magnetically active stars thanks to their sensitive high-resolution grating spectra.

\subsection{The FIP vs inverse FIP effects}

\begin{figure}[!t]
\centering
\resizebox{\hsize}{!}{\includegraphics[angle=-90]{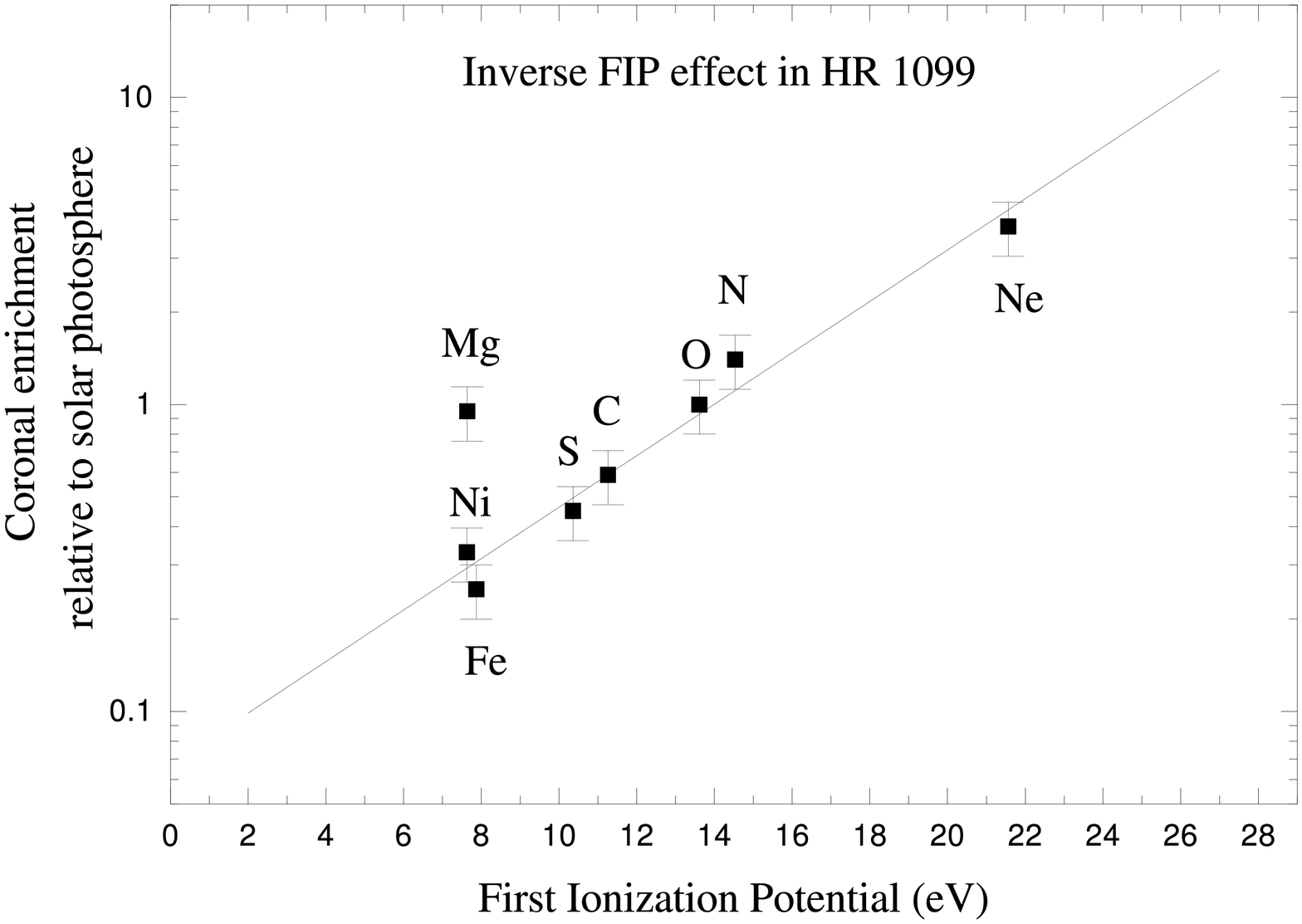}}
\caption{The inverse FIP effect as measured by \protect\cite*{brinkman01} in a deep \xmm\  observation of HR 1099.%
\label{fig:brink01}}
\end{figure}

\begin{figure}[!t]
\centering
\resizebox{\hsize}{!}{\includegraphics[angle=0]{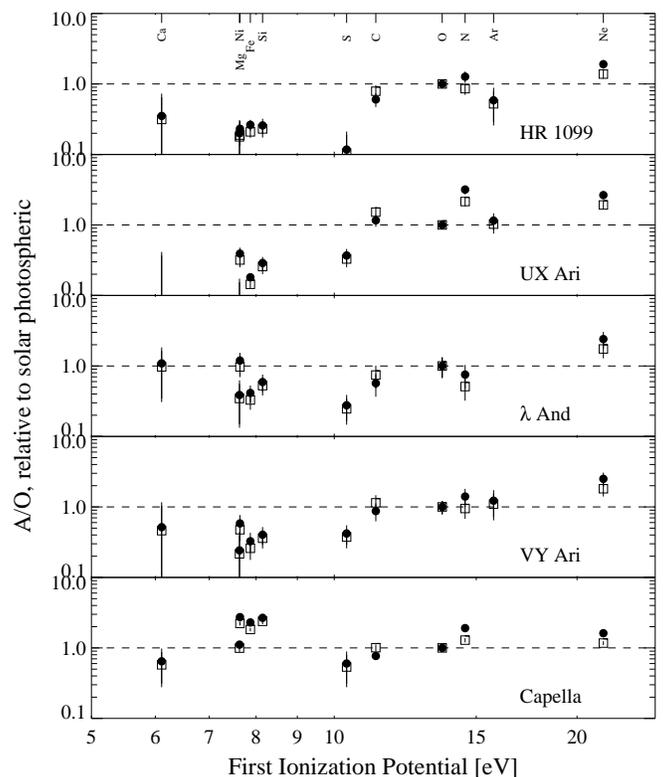}}
\caption{A systematic study of the coronal abundances in RS CVn binaries showed an inverse FIP
effect in the most active stars, whereas the intermediately active Capella showed 
no specific bias (\protect\cite{audard03b}). The symbols refer to different plasma emission codes.
\label{fig:audard03}}
\end{figure}

A new pattern was observed in the active RS CVn binary system HR 1099 in which high-FIP ($>10$~eV) elements
are enhanced relative to the abundances of low-FIP elements
(Fig.~\ref{fig:brink01}; \cite{brinkman01,drake01,audard01a}). 
A detailed analysis of several active RS CVn binaries showed a similar 
inverse FIP effect (Fig.~\ref{fig:audard03}; \cite{audard03b}). 
Several studies with \chan\  and \xmm\  have confirmed the depletion of low-FIP 
elements (e.g., Fe), and the relative enhancement of high-FIP elements (e.g., noble gases Ar and 
Ne) in magnetically active stars (e.g., \cite{guedel01a,guedel01b,huenemoerder01,%
gondoin02,raassen02,stelzer02,%
drake03a,gondoin03a,gondoin03b,huenemoerder03,osten03a,sanz03,vdbesselaar03,%
argiroffi04,audard04,gondoin04,maggio04,schmitt04%
})

 In solar analogs of different activity levels and ages, but of photospheric
composition similar to the Sun's, the FIP bias is more complex: whereas young, active solar proxies showed an
inverse FIP effect similar to that in the active RS CVn binaries, old, inactive stars show a solar-like FIP
effect, suggesting a possible transition from one FIP pattern to the other with the
level of activity (\cite{guedel02a,telleschi05}). 
The lack of significant FIP bias in Capella (\cite{audard01b,audard03b,argiroffi03}), a wide RS CVn
binary of interemediate activity, suggests that a transition in FIP bias occurs in this class of 
magnetically active stars as well. A solar-like FIP effect is also found in 
several giants (\cite{scelsi04,garcia05}).

Although encouraging, the above general picture faces a significant challenge with Procyon. 
Whereas the cool, inactive coronae of $\alpha$ Cen AB both show solar-like FIP effects 
(\cite{raassen03a}), no FIP bias is measured in the F-type 
subgiant (\cite{raassen02,sanz04}), consistently with \euve\ 
(\cite{drake95}). The FIP bias transition scenario, however, requires a strong solar-like FIP effect in
cool, inactive stars.

\begin{figure}[!t]
\centering
\resizebox{0.95\hsize}{!}{\includegraphics[angle=0]{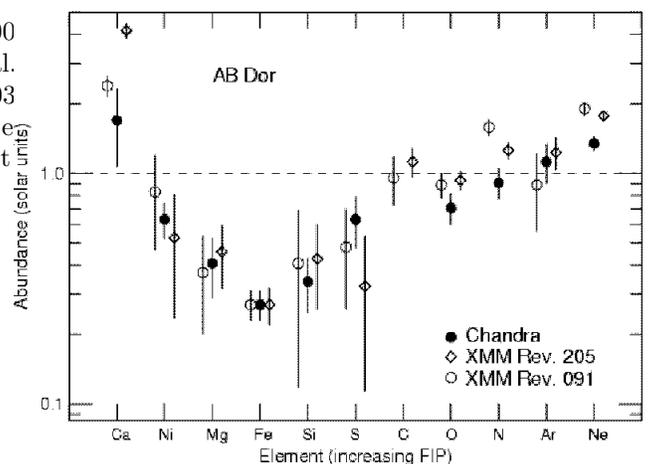}}
\caption{The coronal abundances in the fast rotating AB Dor. Notice the increase abundances at very low FIP
(Ca and Ni; \protect\cite{sanz03}).
\label{fig:sanz03}}
\end{figure}

\begin{figure*}[!t]
\centering
\resizebox{0.475\hsize}{!}{\includegraphics[angle=0]{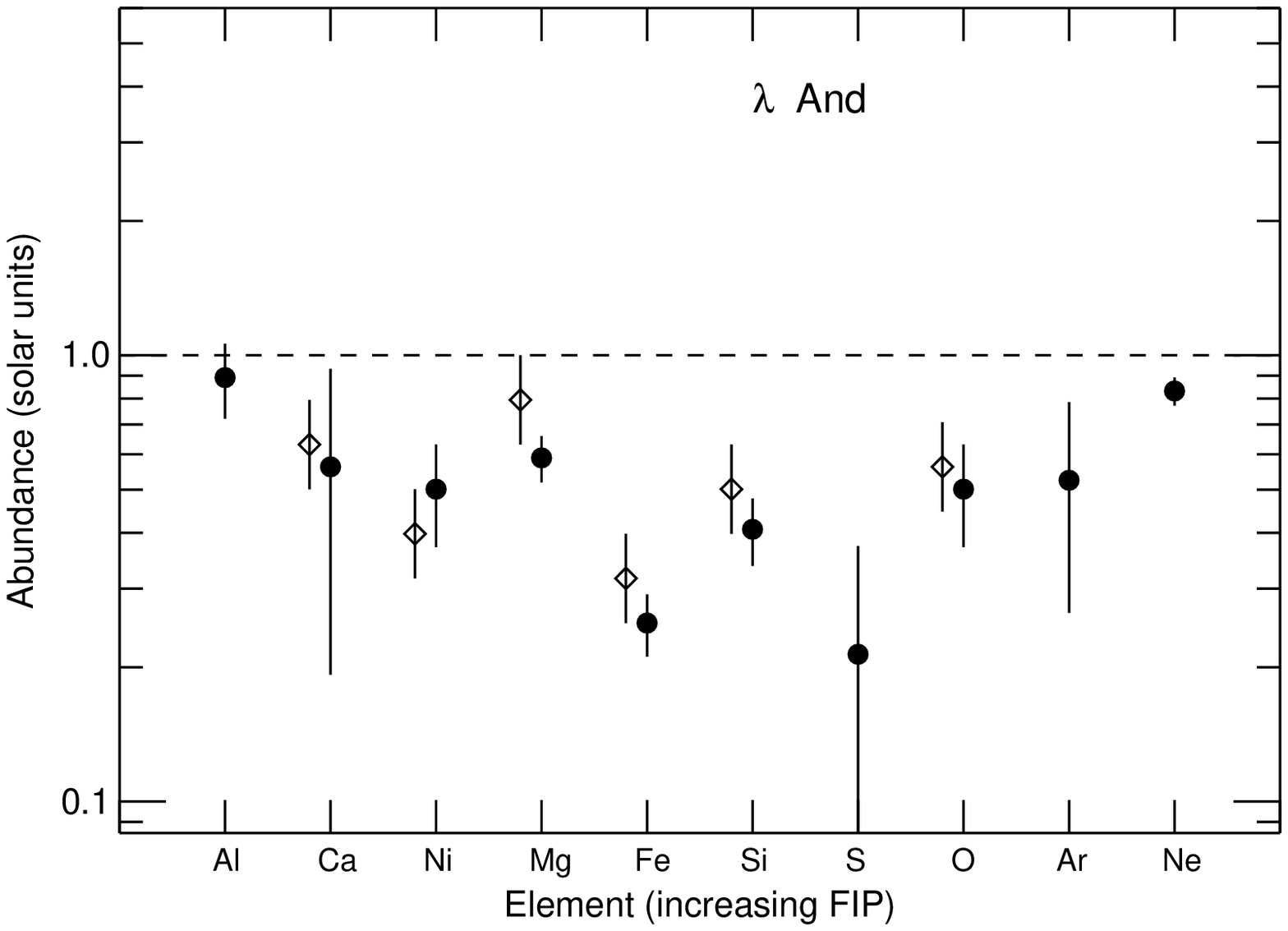}}
\resizebox{0.475\hsize}{!}{\includegraphics[angle=0]{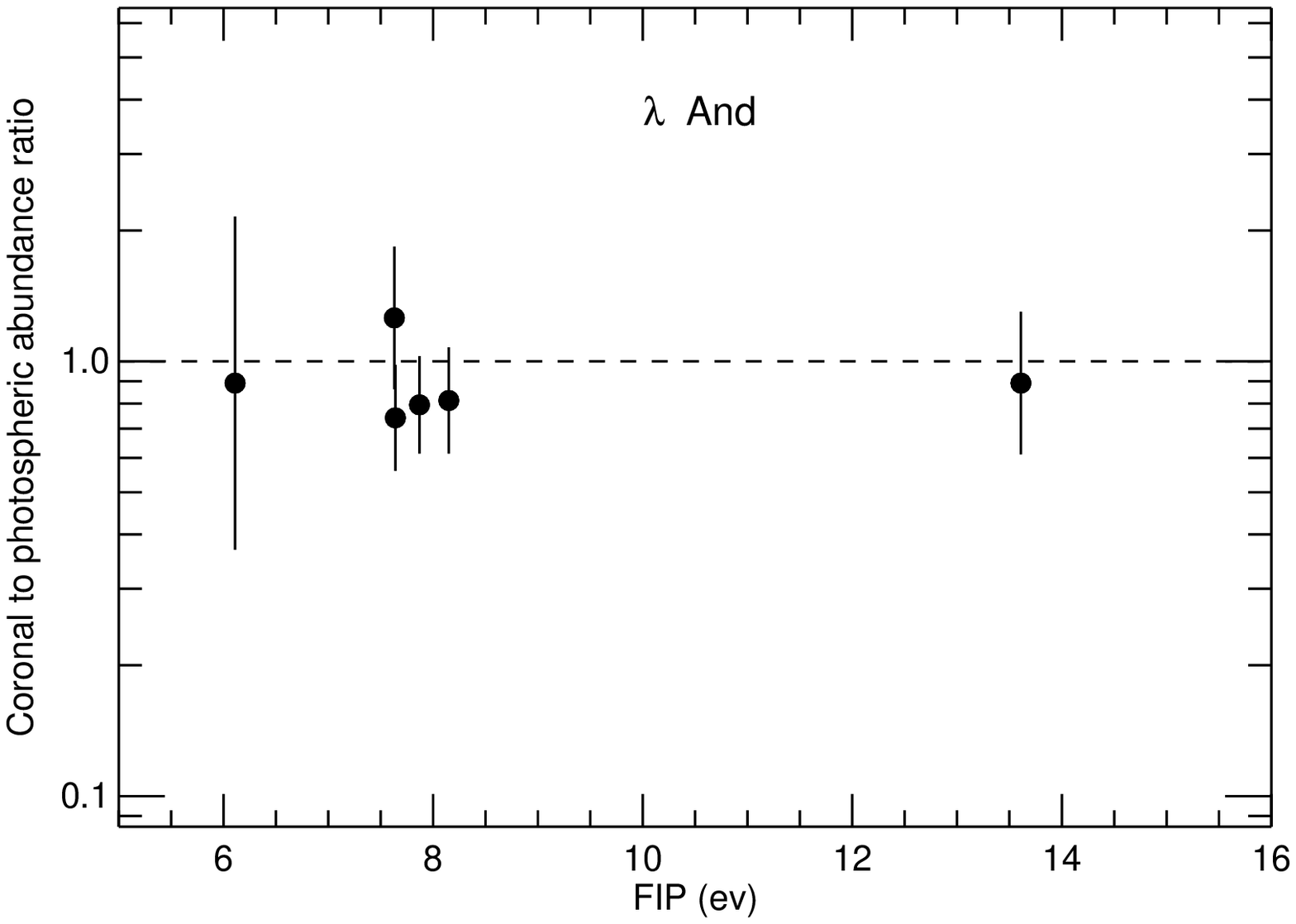}}
\caption{Coronal abundances in the RS CVn binary $\lambda$ And. The U-shape pattern can be seen when the solar
photospheric set is used as comparison (Left). On the other hand, no FIP bias is visible when stellar
photospheric  abundances are used (adapted from \protect\cite{sanz04}; see also \protect\cite{audard03b}).%
\label{fig:sanz04}}
\end{figure*}

Furthermore, several studies have reported that, whereas coronal abundances decrease with decreasing FIP in 
active stars, a turnover occurs at very low FIP where the abundances of the respective elements
steeply increase (Fig.~\ref{fig:sanz03};  \cite{sanz03,osten03a,huenemoerder03,argiroffi04}). 
The explanation for such a turnover remains unclear, as different
analysis techniques provide different results for the same data sets (e.g., \cite{argiroffi04,telleschi05}).
It is worthwhile to mention that, in the Sun's corona, the abundances of elements at very low FIP
are often larger than that of the low-FIP (Fe, Mg, Si) elements (see \cite{feldman00} for a discussion). 

An important caveat needs, however, to be raised: most studies rely on the solar photospheric abundances as
the standard comparison set (e.g., \cite{anders89,grevesse98}). But the derivation of stellar photospheric abundances in magnetically active 
stars is challenging as such stars rotate fast, thus the signal-to-noise ratio of lines against the continuum 
is significantly reduced. In addition, spots and plages can distort the line shapes as the star rotates.
Consequently, photospheric abundances can be highly uncertain, with major differences found in the literature, or,
as is often the case, they are not available altogether (except perhaps the metallicity [Fe/H]).

The FIP coronal pattern could be different when real photospheric values for the individual stars are taken 
into account (Fig.~\ref{fig:sanz04}; \cite{audard03b,sanz04}). 
However, since the solar analogs sample used by \cite*{guedel02a} and \cite*{telleschi05} have
photospheric compositions which are believed to be similar to the Sun's, the finding of a 
transition from
inverse FIP to solar-like FIP effect with decreasing activity level appears to be robust. Also, several stars
with known Fe photospheric abundances show a marked Fe depletion in the corona (e.g.,
\cite{huenemoerder03,audard04}), suggesting that the inverse FIP effect is real.

Finally, the ill-posed problem of spectral inversion needs to be recalled in this review. \cite*{craig76a} and
\cite*{craig76b} showed that statistical uncertainties can introduce significant
uncertainty in the reconstructed emission measure distribution. In addition, systematic uncertainties
introduced by, e.g., inaccurate calibration or incorrect atomic parameters in databases can increase the scatter
further. Consequently, the effects on the derivation of coronal abundances can be severe unless
external assumptions are introduced. Future studies need to take into account this basic problem, and 
 to compare
the output with different analysis techniques and atomic databases 
(e.g., \cite{audard04,schmitt04,telleschi05}).

\subsubsection{The models}

Models that aim at explaining the solar FIP effect are plenty (see the reviews by \cite{henoux95,henoux98}; 
see also \cite{arge98,schwadron99,mckenzie00}). On the other hand, the recent \chan\  and
\xmm\  results still lack theoretical explanations. \cite*{guedel02a} suggested that the non-thermal
electrons seen in magnetically active stars could explain the inverse FIP effect, and possibly the FIP
effect observed in inactive stars. Although promising, the scenario still remains in an infant state.
To my knowledge, the model by \cite*{laming04} is the only comprehensive theoretical model
that specifically addresses the inverse FIP effect in stars.  The model proposes a unified picture of both the
FIP and inverse FIP effects by exploring the effects on the upper chromospheric plasma of the wave
ponderomotive forces. \cite*{laming04} suggested that fine tuning of the model parameters could turn a solar FIP effect
into an inverse FIP effect. Further theoretical works are needed to assess the validity of the proposed model,
and to develop new models and ideas. The work in stellar coronae can thus provide important ideas
and constraints on the solar element fractionation models.

\subsection{Signatures of the CNO cycle}
\label{sec:cno}

\begin{figure}[!t]
\centering
\resizebox{\hsize}{!}{\includegraphics[angle=0]{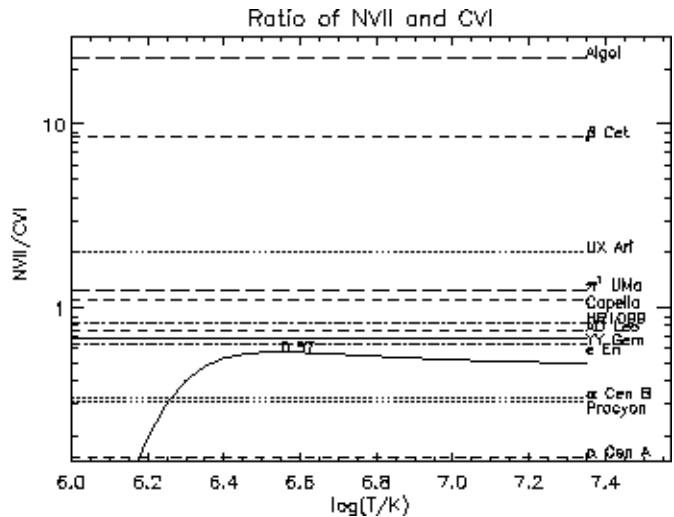}}
\caption{Line ratios of {\rm N~\textsc{vii}} and {\rm C~\textsc{vi}} shown with the theoretical line ratio
for solar photospheric abundances and the MEKAL atomic database as a function of the plasma temperature. 
The line ratios in Algol and $\beta$ Cet lie significantly above the solar ratio, indicating the presence
of CNO-cycle processed material at the stellar surface (\protect\cite{schmitt02}).%
\label{fig:schmitt02}}
\end{figure}

\begin{figure}[!t]
\centering
\resizebox{\hsize}{!}{\includegraphics[angle=0]{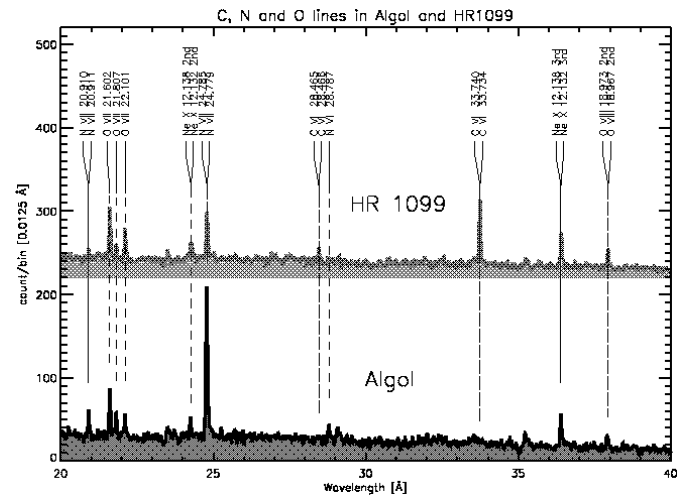}}
\caption{Extracts of the LETGS spectra of  HR 1099 (top) and Algol (bottom). Notice the
enhanced {\rm N~\textsc{vii} Ly$\alpha$} line at $24.7$~\AA\  and the faint  {\rm N~\textsc{vii}
Ly$\alpha$} line in Algol, whereas the situation is almost reversed in HR 1099 (\protect\cite{drake03a}).
\label{fig:drake03}}
\end{figure}

Whereas most stars show coronal abundances affected by element fractionation in their upper atmospheres, a few
stars have shown enhanced N emission lines compared to what could be expected by, say, an inverse FIP effect.
In particular, the N~\textsc{vii}/C~\textsc{vi} line ratio, which is about constant for 
temperatures of above 2.5~MK, turns out to be significantly non-solar in several
stars  (Fig.~\ref{fig:schmitt02}). Such high line ratios were interpreted as a signature of the CNO-cycle 
processed material (\cite{schmitt02}). Using the N/C abundance ratio as a diagnostic tool, 
\cite*{drake03a} showed that the secondary star in Algol must have lost at least half of its 
initial mass onto the primary through accretion (Fig.~\ref{fig:drake03}). 
Evidence of the CNO cycle has also been found in the Algol-type RZ Cas (\cite{audard05}). 
Some giant stars show evidence for CNO-processed material at
their surface as well ($\beta$ Cet: \cite{schmitt02}; YY Men: \cite{audard04}). Finally, \cite*{drake03b}
measured a strong [C/N] depletion in the pre-cataclysmic binary V471 Tau which they interpreted as
observational evidence of the common envelope phase of the system.

\begin{figure}[!b]
\centering
\resizebox{\hsize}{!}{\includegraphics[angle=0]{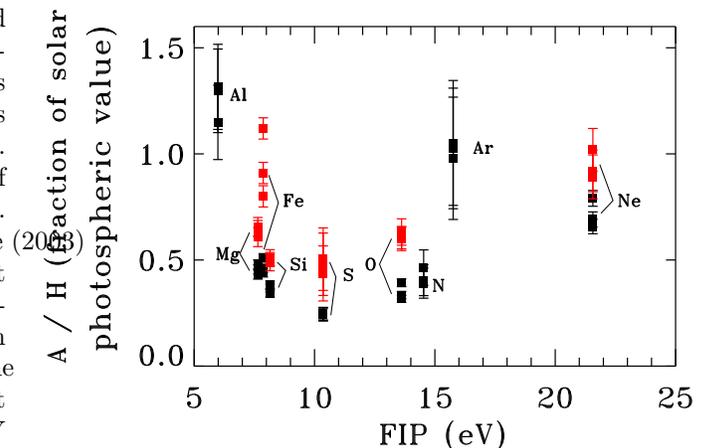}}
\caption{Coronal abundances in $\sigma^2$ CrB. During a stellar flare, all element abundances increase, with
no specific FIP-related bias (\protect\cite{osten03a}).%
\label{fig:osten03a}}
\end{figure}

\subsection{Abundance variations during stellar flares}
In the chromospheric evaporation model, fresh material is brought up from the chromosphere into the corona
after electron beams impact and heat up the lower parts of the atmosphere (\cite{antonucci84}).
Therefore, X-ray spectra are expected to reflect abundances closer to the photospheric (actually
chromospheric) composition. The increase of the average metallicity, $Z$, has been measured in several stellar
flares with previous missions. \asca\  measured in a flare in UX Ari a selective increase of low-FIP elements 
whereas high-FIP element abundances stayed nearly constant (\cite{guedel99,osten00}). A similar behavior was
observed with gratings in other stars (\cite{audard01a,raassen03b}). However, this picture is not unique: 
other flare spectra displayed no FIP-related bias, i.e., all element abundances increased (Fig.~\ref{fig:osten03a}; 
\cite{osten03a,guedel04b}). In addition, some flares show no evidence of abundance increase at 
all (\cite{huenemoerder01,vdbesselaar03}). In conclusion, the abundance behavior during flares
remains unclear, possibly due to the low signal-to-noise ratios of time-dependent flare spectra.

\section{Stellar Flares}
\label{sec:flares}

Magnetically active stars have been known to display numerous flares in general (with the exception of a few
notable examples, e.g., Capella and Procyon). Such flares can show extreme X-ray luminosities (up to a few
percents of the stellar bolometric luminosity), very hot temperatures of more than 10~MK, and can last from
minutes up to several days. In addition, flares can bring ``fresh'' chromospheric material up into the corona
(\cite{antonucci84}). Recent works also suggest that flares can act efficiently as stochastic agents to heat the
corona in the Sun (\cite{parker88}) and in stars (\cite{guedel97,audard00,kashyap02,guedel03a,arzner04}). 
\chan\  and \xmm\  have observed several flares in magnetically active stars, including rather
bright events.
I summarize below the results obtained with those satellites and in multi-wavelength campaigns.

\begin{figure}[!t]
\centering
\resizebox{\hsize}{!}{\includegraphics[angle=0]{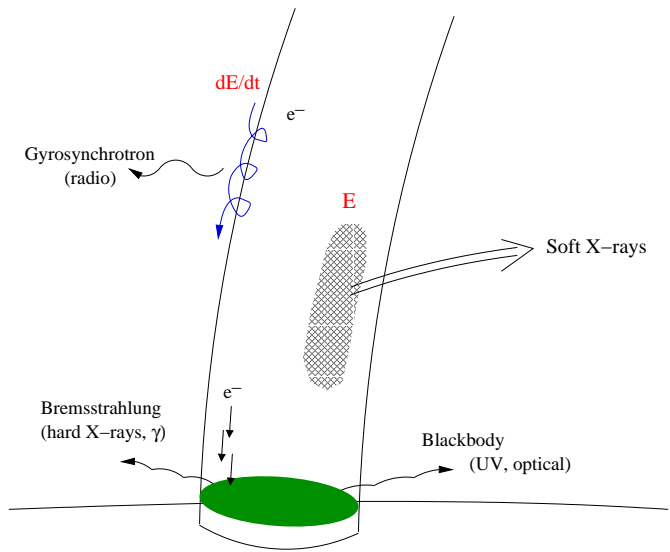}}
\caption{Sketch of the chromospheric evaporation model and the Neupert effect (\protect\cite{neupert68}; 
see text for details).%
\label{fig:neupert}}
\end{figure}

\subsection{The Neupert Effect}

The Neupert effect (\cite{neupert68}) describes the correlation between the microwave/radio light curve of flares
and their X-ray light curve. The latter follows relatively closely the time integral of the former, and this effect has
been explained in the context of chromospheric evaporation (Fig.~\ref{fig:neupert}; \cite{antonucci84,fisher85}). 
A beam of non-thermal electrons is accelerated from the magnetic reconnection site. Part of the electron beam gets trapped 
in the magnetic flux tubes and radiate gyrosynchrotron emission in the radio, and part of it impacts on the dense chromosphere 
producing hard X-ray non-thermal bremsstrahlung (and almost simultaneously broadband optical light). The beam heats up
the upper chromosphere to coronal temperatures; the plasma in turn expands into coronal loops and subsequently cools
radiatively in X-rays. The X-ray emission roughly scales with the thermal energy content deposited
by the electron beam, whereas the radio/hard X-ray/optical emission relates to the energy rate (since these emissions are
spontaneous, i.e., the radiative time scales are much shorter than in the X-rays).

\begin{figure}[!t]
\centering\hspace*{5mm}
\resizebox{0.95\hsize}{!}{\includegraphics[angle=0]{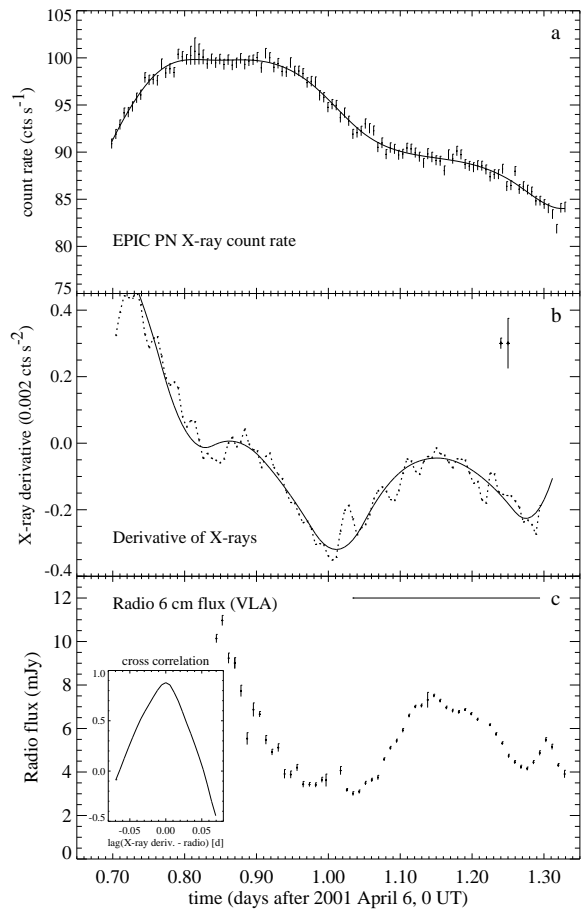}}
\caption{Evidence for the Neupert effect in the RS CVn $\sigma$ Gem. Shown are the light curves observed in
X-rays (top), its X-ray derivative (middle), and in radio (bottom). The inset shows a cross-correlation 
function between the X-ray derivative and radio light curves (\protect\cite{guedel02b}).%
\label{fig:guedel02siggem}}
\end{figure}

Evidence of the Neupert effect has been found in the RS CVn binary $\sigma$ Gem (Fig.~\ref{fig:guedel02siggem}; 
\cite{guedel02b}). Whereas the VLA radio  and \xmm\  pn X-ray light curves show no apparent correlation,
the derivative of the X-ray light curve follows closely the radio, as expected from the Neupert effect. 
A rough assessement of the flare energy budget showed that the energy in the non-thermal electrons is probably
sufficient to heat the observed
plasma. Indeed, under realistic assumptions, \cite*{guedel02b} obtained an injected energy $E \sim 10^{33
- 36}$~ergs, similar to the X-ray emitted $E_\mathrm{X} \sim 4 \times 10^{34}$~ergs. In a systematic study of
the X-ray and radio simultaneous observations of several M-type flare stars, \cite*{smith05} also observed
several cases of a Neupert effects, although they reported events where no radio counterparts were observed to X-ray
flares, and vice-versa. They argued that, like in the Sun, not all flares should display a
Neupert effect since the observable fluxes can originate from different mechanisms (e.g., radio masers) or 
spatial structures.

\begin{figure}[!t]
\centering
\resizebox{\hsize}{!}{\includegraphics[angle=0]{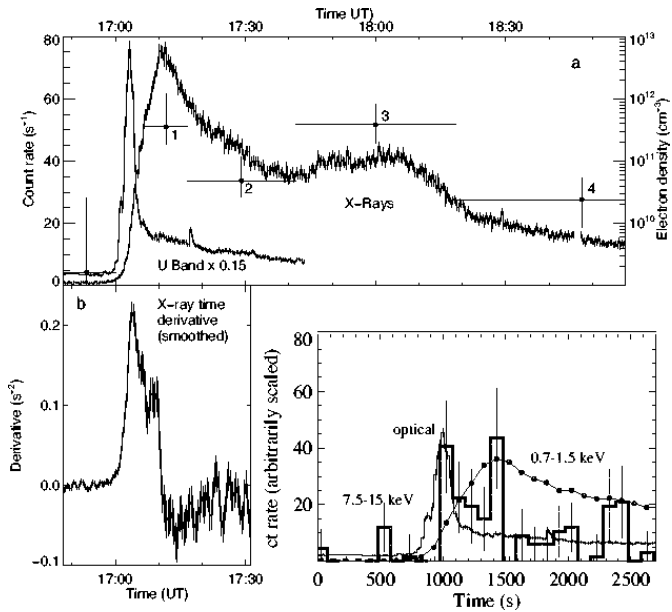}}
\caption{The Neupert effect in Proxima Centauri. The X-ray and optical light curves are shown in the top
panel; the X-ray derivative is shown in the bottom left panel, whereas the hard X-ray ($7.5-15$~keV) light curve is
shown as a histogram in the bottom right panel (adapted from \protect\cite{guedel02c}).%
\label{fig:guedel02proxcen}}
\end{figure}

\begin{figure}[!t]
\centering\hspace*{5mm}
\resizebox{0.98\hsize}{!}{\includegraphics[angle=0]{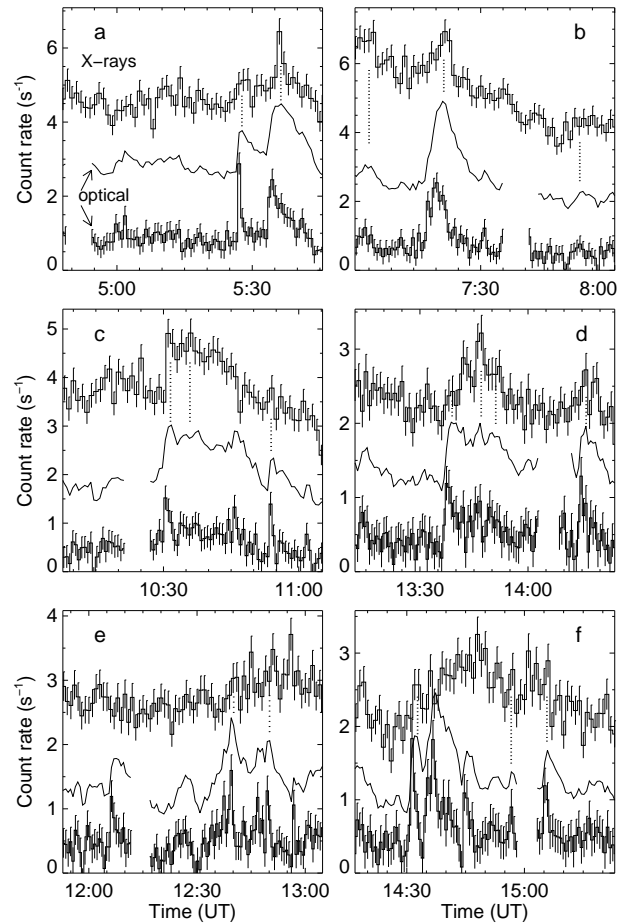}}
\caption{Several small flares are detected in the pseudo-``quiescent'' X-ray and optical light curves of 
Prox Cen. Some also display a Neupert effect (\protect\cite{guedel02c}).%
\label{fig:guedel02proxcen2}}
\end{figure}

A large flare in Proxima Centauri was caught with \xmm. Simultaneous coverage with the Optical Monitor
proved helpful to observe the Neupert effect (\cite{guedel02c}). In analogy with $\sigma$ Gem, the X-ray 
derivative light curve closely matched the optical $U$-band light curve. X-ray photons in the range 
$7.5-15$~keV also occurred in the early phase of the flare (Fig.~\ref{fig:guedel02proxcen}). The peak flare
X-ray luminosity reached $3.9 \times 10^{28}$~\ergps, about hundred times the
pre-flare level ($6 \times 10^{26}$~\ergps), and the total energy radiated in
X-rays was about $1.5 \times 10^{32}$~ergs. There was spectroscopic evidence of density
variations from $< 2 \times 10^{10}$~\cmcc\  (pre-flare) to $10^{11-12}$~\cmcc\ 
at flare peak. Together with the EMs, coronal volumes and masses
could be derived as well. \cite*{reale04} analyzed the Prox Cen flare with
hydrodynamical codes and found that it could be described by an initial flare heating in a single loop ($\sim
10^{10}$~cm) followed by additional heating in an arcade-like structure. \cite*{guedel04b} found a similar
loop size from a 2-ribbon flare modeling.

Further evidence of the Neupert effect and chromospheric evaporation was detected in the pre-flare light curve
in Prox Cen, before the giant flare. The ``quiescent'' light curve turned out to be quite variable: a
multitude of small events were detected in the X-ray \emph{and} optical light curves, with the smallest events
having X-ray luminosities $L_\mathrm{X} \sim 10^{26}$~\ergps, i.e., similar to modest solar flares
(\cite{guedel02c}). Many of them showed a Neupert effect, but some did not which is no different than in the Sun 
(\cite{dennis93}). Such observations demonstrate that what is perceived as quiescence could in 
fact be explained by the superposition of a multitude of small heating flares.

\subsection{The Flare Emission Measure Distribution}

\begin{figure}[!t]
\centering
\resizebox{\hsize}{!}{\includegraphics[angle=0]{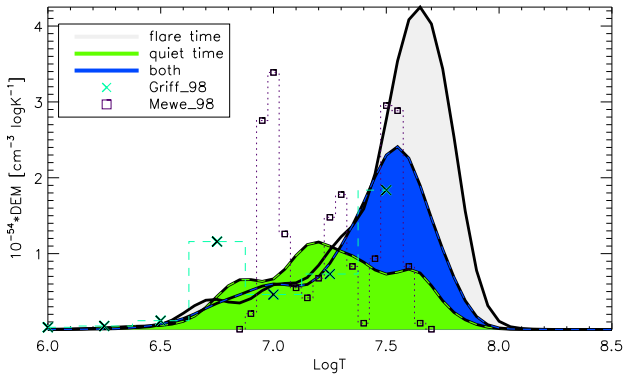}}
\caption{The EMD of the RS CVn binary II Peg (\protect\cite{huenemoerder01}). 
The time-averaged EMD is a composite of the ``quiescent'' EMD and the flare EMD. 
\label{fig:huenemoerder01}}
\end{figure}

\begin{figure}[!b]
\centering
\resizebox{0.89\hsize}{!}{\includegraphics[angle=0]{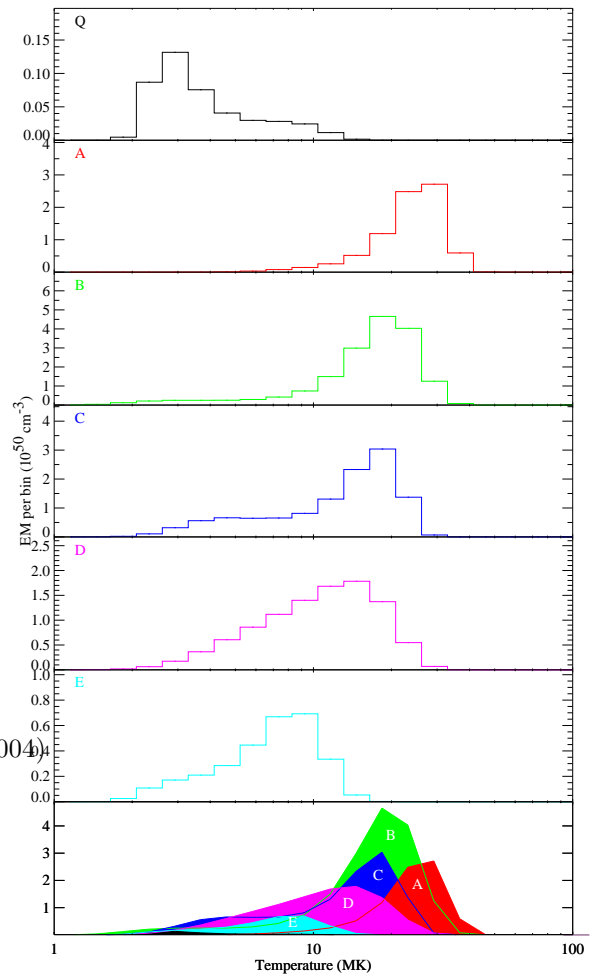}}
\caption{EMDs of Prox Cen (\protect\cite{guedel04b}). The ``quiescent''
EMD is shown at the top, whereas the time-dependent flare EMD is shown in the subsequent
panels. The bottom panel shows a superposition of all EMDs on the same vertical scale.
\label{fig:guedel04}}
\end{figure}

Flare spectra have displayed spectroscopic characteristics of high temperatures, such as a well-developed brems\-strahlung
continuum and emission lines of highly ionized Fe (e.g., \cite{audard01a,osten03a}). The reconstructed
emission measure distribution (EMD) during flares indeed show strong EMs at high
temperatures (Fig.~\ref{fig:huenemoerder01}; \cite{huenemoerder01,raassen03b,vdbesselaar03}). Time-dependent
analyses actually revealed that the flare EMD evolves from a dominant high temperature in the
early phase followed by a gradual decrease in emission measure and temperature
(Fig.~\ref{fig:guedel04}; \cite{guedel04b}).

\begin{figure}[!t]
\centering
\resizebox{\hsize}{!}{\includegraphics[angle=0]{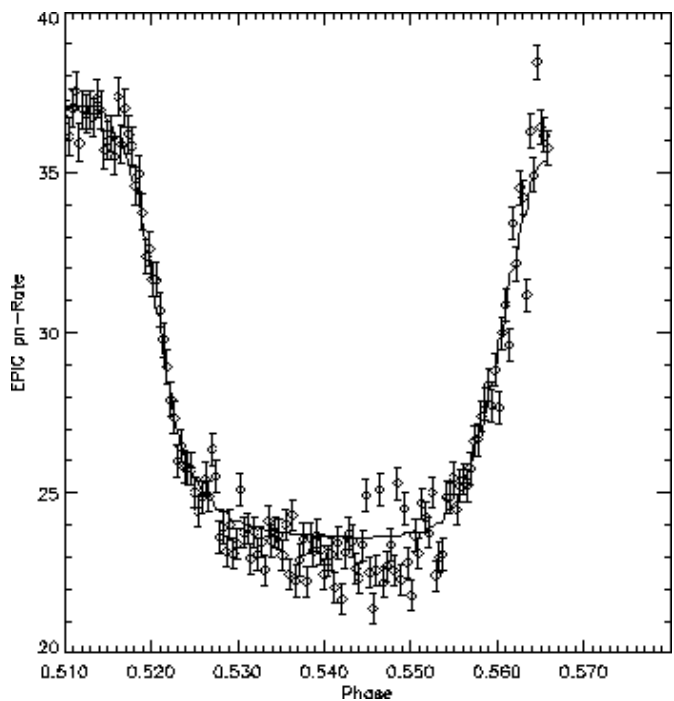}}
\caption{The \xmm\  X-ray light curve of an eclipsed flare in Algol after correction using a flare decay
profile (\protect\cite{schmitt03}). The best-fit model is shown as a solid curve. Fig.~\ref{fig:schmitt03d}
shows the reconstructed spatial image.
\label{fig:schmitt03}}
\end{figure}

Although the EMD temporarily evolves to high temperatures and returns to a pre-flare shape, the question arises
whether flares as a statistical ensemble (i.e., from small to large flares) can shape the EMD of stellar coronae
(e.g., \cite{guedel97,guedel03a}). In this framework, \cite*{guedel04b} noted that the large flare in the modesty
X-ray active Prox Cen was equivalent to a small ``wiggle'' in the extremely X-ray active flare star YY Gem 
and further emphasized that the flare
spectrum of Prox Cen closely matched YY Gem's \emph{quiescent} spectrum. \cite*{audard04} also showed that
the high-temperature tail of YY Men's EMD can be ascribed to numerous small flares despite the lack of obvious 
flares in its light curve.

\subsection{Eclipse Mapping}

\begin{figure}[!t]
\centering
\resizebox{\hsize}{!}{\includegraphics[angle=0]{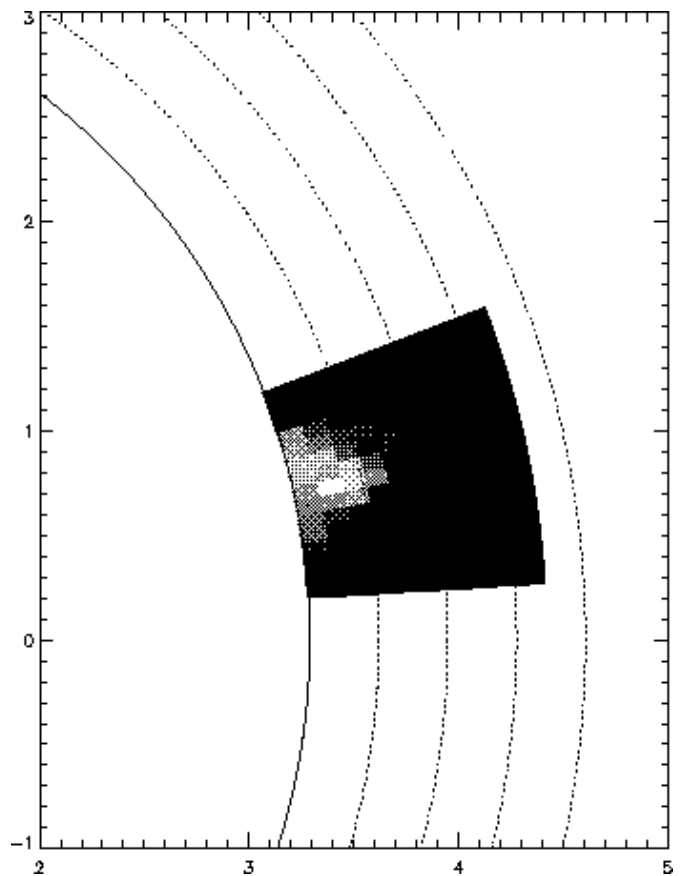}}
\caption{The reconstructed spatial image of the flare in Algol (see Fig.~\ref{fig:schmitt03}). The solid
curve describes the photosphere of the K-type secondary, whereas the dotted curves denote successive
heights in units of $0.1 R_\star$ (\protect\cite{schmitt03}).
\label{fig:schmitt03d}}
\end{figure}

\cite*{schmitt03} observed in Algol an eclipse of a flare that was in progress.
Assuming a flare decay profile, they modeled a rectified light curve
with an eclipse mapping code to derive the properties of the eclipsed flare. \cite*{schmitt03}
constrained the location of the flare to near the limb of the magnetically active K2 star
at a height of $\sim 0.1 R_\star$. The cumulative density distribution indicates densities of
at least $10^{11}$~\cmcc\  and up to $\sim 2 \times 10^{11}$~\cmcc. Similar densities are
derived from spectroscopy (O~\textsc{vii} He-like triplet). A similar
eclipsed flare was observed with \sax\  (\cite{schmitt99,favata99}).

Eclipse mapping is a powerful means to obtain indirect information on the spatial structure of
stellar coronae. \cite*{guedel01b}, \cite*{guedel03b}, and \cite*{guedel05} used
a similar method to obtain spatial maps of the eclipsing binaries YY Gem, $\alpha$
CrB, and CM Dra. Inhomogeneous coronal maps were derived, albeit with characteristics
different from the Sun's (e.g., high altitudes). \cite*{audard05} observed the eclipsing Algol-type RZ Cas with \xmm\
and the VLA. Only shallow eclipses were detected with a different timing in the different
wavelength regimes, suggesting extended emitting sources. Another technique to derive spatial
information includes spectroscopy. \cite*{ayres01} monitored the centroid of
the Ne~\textsc{x} Ly$\alpha$ line in HR 1099 over half an orbital period and concluded that the
bulk of the X-ray emission was confined to the K-type secondary. \cite*{brickhouse01} 
observed line centroid shifts in the contact binary 44i Boo. They found that derived that two active
regions on the primary star at high latitudes could reproduce the observed line profile shifts
and the X-ray light curve. YY Gem showed variable line broadening (\cite{guedel01b}), whereas
line shifts were measured in Algol (\cite{chung04}).

\subsection{Cool flares}

\begin{figure}[!t]
\centering
\resizebox{\hsize}{!}{\includegraphics[angle=0]{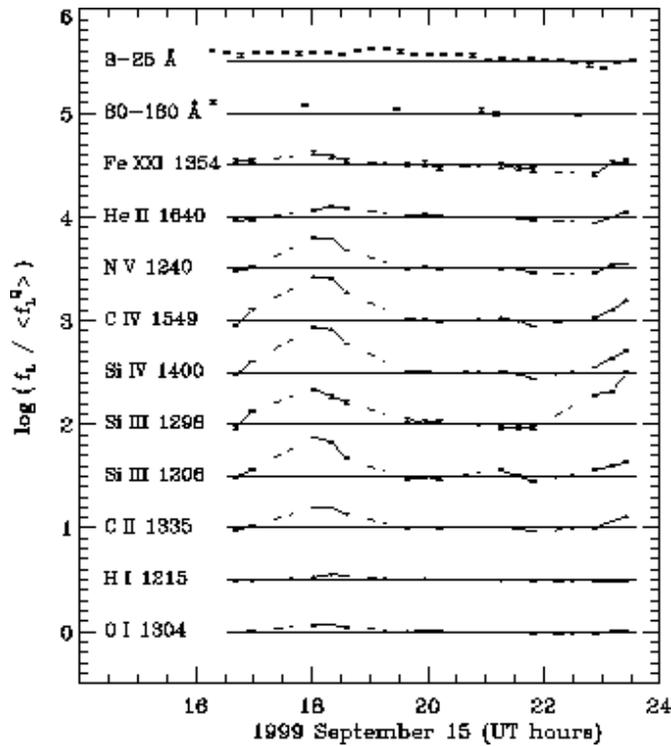}}
\caption{Light curves in X-rays, EUV, and in several UV emission lines (\protect\cite{ayres01}). The first top
three curves are sensitive to hot ($>0.1$~MK) plasma. Notice that the UV ``cool'' flares are not detected in 
those curves.%
\label{fig:ayres01}}
\end{figure}

Common wisdom assumes that flares show typically extreme high temperatures. However, \cite*{ayres01}
observed ``cool'' flares in a multi-wavelength campaign on HR 1099. The flares displayed temperatures less
than 0.1~MK since there was no signal detected in the X-ray or EUV regimes, neither in the Fe~\textsc{xxi}
coronal line at $\lambda 1354$~\AA\  (Fig.~\ref{fig:ayres01}). Such cool flares were interpreted as similar to transition zone
explosive events observed in the Sun (\cite{dere89}). \cite*{osten03b} also reported a flare in radio/optical
with no X-ray counterpart in EV Lac.

\section{Concluding Remarks}
The previous pages have provided a glimpse of the exciting new results in stellar coronal physics
obtained with \xmm\  and \chan\  to date. The wealth of scientific publications five years after the
satellites' launches is a lively proof of rapid advances in our field. I have concentrated on selected aspects,
i.e.,  the elemental composition and stellar 
flares. Further topics are covered in a companion review in these proceedings (\cite{ness05}). Furthermore, numerous
poster papers in these proceedings complement the present review as well. The field of stellar coronae
has matured and profited significantly from \xmm\  and \chan. Nevertheless, future studies are needed
to deepen our knowledge on stellar coronae in X-rays and at other wavelengths, and in particular 
to understand the connection between the Sun and magnetically active stars.

\begin{acknowledgements}

This review is dedicated to the late Dr. Rolf Mewe whom I met while a graduate student, and who left us
too early. His knowledge, his kindness, and his friendship will be greatly missed.

I would like to thank the SOC of the CS13 workshop for their invitation to present this review. I
also acknowledge fruitful discussions with colleagues over the past few years in this research
field. I am grateful to Manuel G\"udel and Rachel Osten for their useful suggestions to improve this
review. Rachel Osten, David Huenemoerder, and Jeremy Drake are also thanked for 
providing electronic versions of figures. Finally, the respective authors, the 
American Astronomical Society, and the editors of Astronomy \& Astrophysics 
are thanked for granting permission to use previously published figures in this review.
\end{acknowledgements}

\end{document}